\documentclass[%
 reprint,
 superscriptaddress,
 amsmath,amssymb,
 aps,
 prl,
]{revtex4-2}

\usepackage{physics}
\usepackage{graphicx}
\usepackage{dcolumn}
\usepackage{bm}
\usepackage{booktabs} 

\usepackage{xcolor}

\begin{document}

\title{High-throughput screening and mechanistic insights into solid acid proton conductors}

\author{Jonas H{\"a}nseroth}
\email{jonas.haenseroth@tu-ilmenau.de}
\affiliation{Institute of Physics and Institute of Micro- and Nanotechnologies, Technische Universit\"at Ilmenau, 98693 Ilmenau, Germany}

\author{Max Gro{\ss}mann}
\affiliation{Institute of Physics and Institute of Micro- and Nanotechnologies, Technische Universit\"at Ilmenau, 98693 Ilmenau, Germany}

\author{Malte Grunert}
\affiliation{Institute of Physics and Institute of Micro- and Nanotechnologies, Technische Universit\"at Ilmenau, 98693 Ilmenau, Germany}

\author{Erich Runge}
\affiliation{Institute of Physics and Institute of Micro- and Nanotechnologies, Technische Universit\"at Ilmenau, 98693 Ilmenau, Germany}

\author{Christian Dre{\ss}ler}
\affiliation{Institute of Physics and Institute of Micro- and Nanotechnologies, Technische Universit\"at Ilmenau, 98693 Ilmenau, Germany}

\date{\today}

\begin{abstract}

Proton-conducting solid acids could enable water-free operation of high-temperature fuel cells.
However, systematic materials screening has, hitherto, been computationally prohibitive. 
Here, we introduce a two-stage high-throughput screening strategy that directly computes proton diffusion coefficients, enabled by machine-learned interatomic potentials fine-tuned to ab initio data.
Starting from more than six million materials, our screening---based on structural motifs rather than empirical descriptors---identifies $27$ high-performing proton conductors, including over ten previously unexplored compounds. 
These include sustainable and commercially available materials, candidates that have not yet been synthesized, organic systems that fall outside conventional design rules, and known proton conductors that validate our approach. 
Importantly, our findings reveal a universal oxygen--oxygen distance of approximately $2.5$~{\AA} at the moment of proton transfer across diverse chemistries, providing mechanistic insight and showing that macroscopic proton conductivity emerges from the interplay between anion rotational dynamics, hydrogen-bond network connectivity, and proton-transfer probability.
\end{abstract}

\maketitle

Proton-conducting membranes in today's fuel cells carry a fundamental constraint: they operate optimally only when sufficiently hydrated \cite{ochi_09, wan23, patel16, prabhakaran12}.
This water dependency is more than just a practical inconvenience, as it restricts operating temperatures and accelerates degradation under dry conditions \cite{steele2001materials, brandon2003recent}. 
Higher operating temperatures in turn would be transformative: they would boost reaction kinetics, streamline thermal management, and strengthen overall system durability.
However, overcoming this central limitation of current proton-conducting membranes requires materials that maintain high conductivity without relying on bulk water \cite{haile2001, boysenand2003high, boysen04, chisholm2005superprotonic, uda_2005, chisholm09, goni12, kim2015characterization, wang2025revealing, grunert2025}.

Solid acids built on oxoanion frameworks have emerged as leading candidates for this challenge.
Unlike conventional hydrated polymers, these materials conduct protons through a mechanism that operates without solvating water.
Instead, protons migrate between neighboring oxoanions via Grotthuss-like hopping---i.e., concerted proton transfer along hydrogen-bond networks accompanied by continual breaking and re-forming of oxygen-hydrogen bonds---coupled to the dynamic reorientation of the anion sublattice \cite{grotthuss, agmon1995grotthuss}.
This fundamentally different transport mechanism points to two essential ingredients for high proton conductivity: rapid proton transfer between adjacent oxoanions and concerted rotational motion that transiently aligns hydrogen-bond networks \cite{baranov03, yamada04, SEVIL20041659, imidazol_polymer, haile07, dressler2020effect, jinnouchi2022}.

These processes are tightly interwoven at the atomic scale, governed by electrostatics, lattice flexibility, and anharmonic dynamics.
Therefore, a predictive understanding demands long-time atomistic simulations that explicitly capture rare proton-hopping events, anion dynamics, and statistically converged diffusion coefficients.

Molecular dynamics (MD) simulations are uniquely suited to describe proton transport in solid acids, as they explicitly resolve proton transfer events and the coupled reorientation of oxoanions.
However, a critical bottleneck remains: empirical force fields cannot capture proton hopping because it involves bond breaking and formation, which this approach cannot model.
Thus, traditional quantitative studies have relied on ab initio molecular dynamics (AIMD), where forces are computed using density functional theory (DFT) \cite{dressler2020effect, dressler2023coexistence}.
While AIMD accurately describes both proton diffusion and anion dynamics, its high computational cost limits simulations to short timescales and small material sets, making systematic screening impractical \cite{marx2000ab, tuckerman2002ab}.

Until recently, this computational barrier meant that most studies screening materials databases for solid acids relied on experimental intuition or static descriptors.
Accordingly, strategies inspired by experiments have explored combinations of oxoanions, cations, and substitutions to merge fast proton transfer with rapid anion rotation, as exemplified by work on CsH$_2$PO$_4$- and CsHSO$_4$-based systems \cite{haile1995superprotonic, haile1997superprotonic, ikeda2014phase, dressler2016, dressler2020a, dressler2020b, dressler2020effect, dressler2023coexistence, wang2022phase, grunert2025, wang2025revealing, haile26}.

Recent advances in machine-learning interatomic potentials (MLIPs) remove this bottleneck by enabling MD simulations with near ab initio accuracy and orders-of-magnitude reduced computational cost \cite{behler2007, bartok2010, drautz2019, friederich2021, batzner2022, reiser2022, mace_1, mace_2}.
Modern, broadly trained MLIPs, often referred to as universal or foundation models \cite{mace_mp, mattersim, kovacs2023, sol3r}, can accurately approximate DFT energies and forces across diverse chemical environments \cite{matbench}. 
This capability enables the high-throughput screening of solid acid proton conductors based on explicit transport descriptors rather than proxies.
However, transport processes probe rare events and configuration-space regions that may be insufficiently represented in the original training data, limiting the quantitative reliability of universal models.
This limitation can be systematically addressed by fine-tuning the models on additional system-specific AIMD data, recovering ab initio accuracy while retaining the efficiency required for long, statistically converged simulations \cite{haenseroth2025amaceingtoolkit, hanseroth2025optimizing, flototto2025large, radova2025fine, liu2025fine, qaisrani2025bridging}. 

Enabled by these methodological advances, {\v{Z}}gun et al.~\cite{vzguns2024uncovering} demonstrated that physically motivated descriptors---identified from AIMD and phonon analysis---can be used to preselect promising candidates from materials databases, followed by machine-learning-accelerated MD to evaluate proton diffusion coefficients for a reduced set of compounds.

Here, we extend the scope and predictive power of this screening strategy.
In this study, we search for promising solid acid materials by directly computing proton diffusion and anion rotational dynamics for solid acid-like compounds drawn from the \textsc{Materials Project} \cite{mp_1, mp_2} (around 200,000 materials) and the \textsc{Alexandria Materials Database} \cite{alexandria, cavignac2025ai} (around $5.8$ million materials), comprising over six million entries in total.
From this vast materials space, candidate structures are first identified based on structural motifs known to facilitate the Grotthuss mechanism in solid acid proton conductors.
The computationally efficient foundation MLIP \textsc{MatterSim} \cite{mattersim} is then employed to enable large-scale screening and ranking of these candidates via MD simulations.
Subsequently, AIMD training data are generated for the most promising materials and used to fine-tune material-specific \textsc{MACE} MLIPs \cite{mace_1, mace_2, mace_mp}, which are employed to obtain quantitatively reliable proton diffusion coefficients and anion rotational dynamics.
By explicitly simulating proton transport rather than relying on chemical intuition or proxy descriptors, this approach enables systematic exploration of solid-acid design spaces and identification of materials with superior proton conductivity in unexplored structural classes and chemical environments.

\section*{Results}

\subsection*{Material selection}

\begin{figure*}
    \centering
    \includegraphics{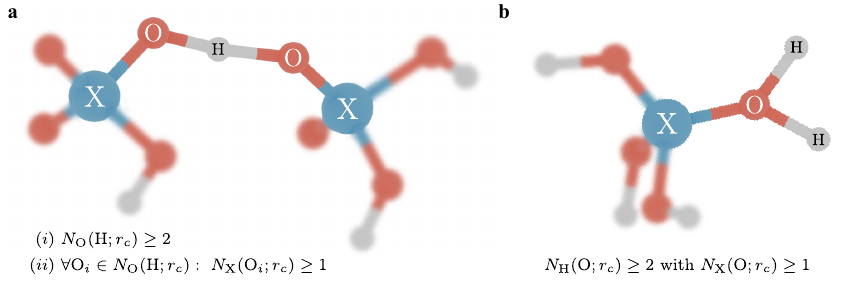}
    \caption{
    \textbf{Structural solid acid motifs used for screening materials databases.}
    Schematic representation of the two complementary proton coordination motifs derived from the solid acids \textbf{a}  CsH$_2$PO$_4$ and \textbf{b}  Cs$_7$(H$_4$PO$_4$)(H$_2$PO$_4$)$_8$ \cite{struct_cdp, struct_cpp}.
    $N_\mathrm{A}(\mathrm{B};r_c)$ denotes the number of atoms of type A around an atom of type B within a radial cutoff $r_c$.
    }
    \label{fig:fig1}
\end{figure*}

Our screening strategy begins with a pool of over six million materials drawn from two databases, the \textsc{Materials Project} \cite{mp_1, mp_2} and the \textsc{Alexandria} database \cite{alexandria, cavignac2025ai}, which were screened for structural motifs characteristic of solid acid compounds, illustrated in Fig.~\ref{fig:fig1}.
This scale allows us to examine a substantially larger candidate pool than the earlier high-throughput solid acid search, thereby broadening the accessible design space and increasing chemical diversity \cite{vzguns2024uncovering}.
To identify candidates with genuine potential as proton conductors, we employ two complementary structural motifs derived from well-characterized solid acids, specifically CsH$_2$PO$_4$ and Cs$_7$(H$_4$PO$_4$)(H$_2$PO$_4$)$_8$ (see Fig.~\ref{fig:fig1}).
The first screening motif identifies materials in which a hydrogen (H) atom is coordinated by two or more oxygen (O) atoms (i.e., $N_\mathrm{O}(\mathrm{H};r_c)\geq2$, where coordination is defined by atoms within a radial distance $r_c$), and each of these oxygen atoms is bonded to an atom X, with X being P, S, Se, or As (see Fig.~\ref{fig:fig1}a).
The X atom is the central atom of the oxoanion group, and because most known solid acids are built from oxoanions with central atoms from groups 15 and 16 of the periodic table, X is restricted to this subset.
This motif represents a proton shared between multiple oxoanion groups, which is a common structural feature of solid acids.
The second screening motif identifies materials in which an oxygen atom bonded to an atom X has two or more hydrogen neighbors ($N_\mathrm{H}(\mathrm{O};r_c)\geq2$), again with X being P, S, Se, or As (see Fig.~\ref{fig:fig1}b).
This motif captures proton-rich oxygen environments that can support extended hydrogen bond networks and facilitate proton transfer.
For both motifs, the identification of oxygen neighbors around hydrogen atoms and the detection of X--O bonds were performed using a radial cutoff $r_c$ of $2.0$~{\AA}.
Materials satisfying either criterion are retained for further analysis.
Note that we impose no further constraints on the structure of materials, allowing us to explore an unusually broad chemical space, including unconventional cationic compositions and sublattices, while focusing exclusively on the proton-conducting oxoanionic framework.

This structural filtering constitutes the primary reduction step of the workflow and eliminates the vast majority of database entries from further consideration.
After filtering for the two motifs, 3,967 compounds remained (2,817 selected by the first motif and 1,150 by the second), constituting a 1,500-fold reduction from the initial material pool.
For each remaining material, supercells containing at least $512$ atoms were constructed to enable statistically meaningful MD simulations (see Supplementary Note 2).

\subsection*{High-throughput screening}

Next, MD simulations with a duration of $100$~ps at $400$~K were performed for all $3,967$ candidates using the foundation MLIP \textsc{MatterSim} \cite{mattersim}.
This stage was used to assess the dynamical stability of each candidate by verifying structural integrity under finite-temperature conditions and identifying persistent proton-bond networks characteristic of solid acids.
In addition, the relative proton transport performance of dynamically stable materials was ranked for subsequent higher-accuracy simulations.
Materials that decomposed into molecular hydrogen or water during the simulation were excluded from further analysis (see Supplementary Note 3). 
The simulation details are described in the Methods section.

The rotational dynamics of the polyanion framework are known to be a key contributor to proton-transfer mechanisms in solid acids \cite{dressler2020effect, wang2025revealing}.
To further refine the candidate set, we quantify anion reorientation by computing the time autocorrelation of X--O bond vectors (see Methods), which tracks the persistence of the X--O bond orientation over time (values near $1$ indicate little reorientation, whereas values approaching 0 indicate substantial rotation) \cite{dressler2020effect}.
We exclude systems with negligible anion dynamics, operationally defined as those maintaining X--O bond-vector autocorrelation coefficients exceeding $0.9$ over the $0$--$100$~ps simulation window (a pragmatic threshold indicating largely preserved bond orientations and only limited reorientation), since efficient proton transport requires fast anion rotation dynamics.
We also incorporated thermodynamic stability constraints, excluding materials with formation energies more than $0.05$~eV above the convex hull.
After applying all filters, $279$ candidates remained after the initial MD simulations with the \textsc{MatterSim} \cite{mattersim} foundation model, corresponding to a manageable subset for ab initio refinement while retaining substantial chemical diversity.

\begin{figure*}[ht]
    \centering
    \includegraphics{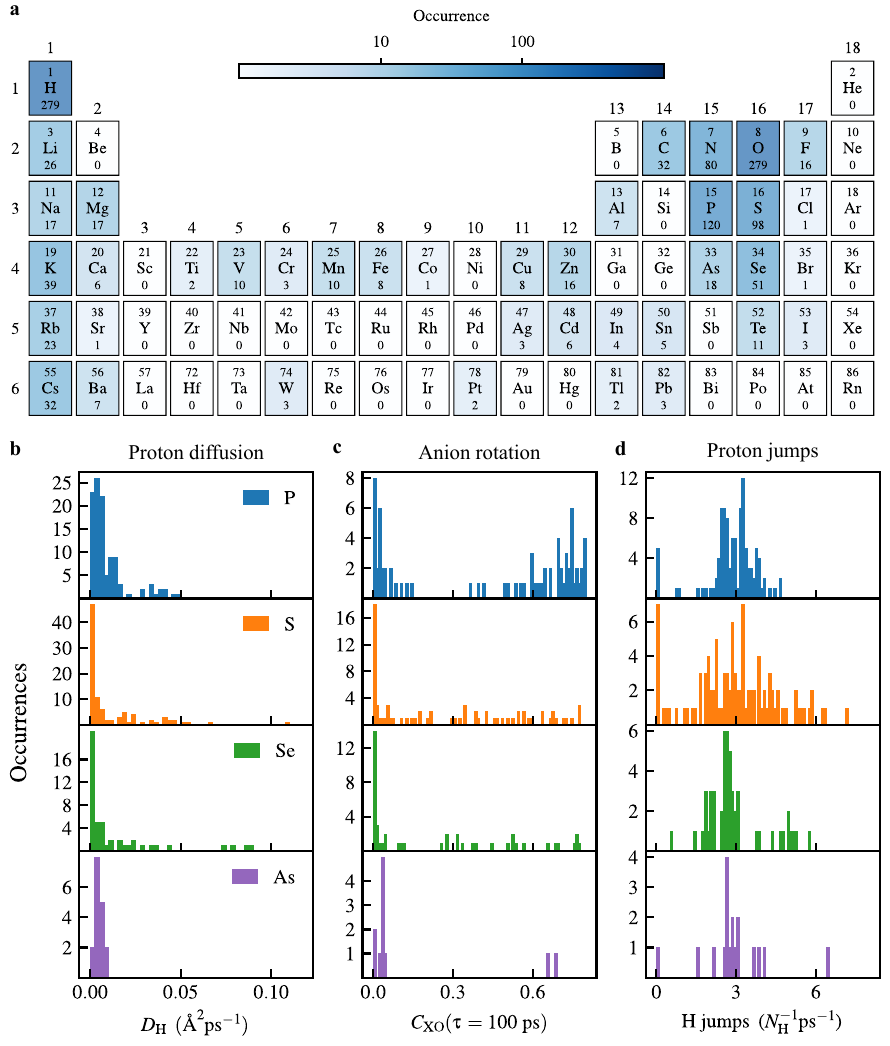}
    \caption{
    \textbf{Overview of the high-throughput screening results obtained using the \textsc{MatterSim} foundation model.}
    Parameters for the MD simulations are given in the Methods section.
    \textbf{a} Elemental composition of the filtered dataset. 
     Colors encode how often each element appears among the filtered structures, with the exact counts printed below the corresponding element symbols.
    \textbf{b}--\textbf{d} Dynamical descriptors extracted from $100$~ps MD simulations (see Methods).
    The proton diffusion coefficients $D_\mathrm{H}$ (\textbf{b}), the X--O bond vector autocorrelation values $C_\mathrm{XO}$ at $\tau = 100$~ps (\textbf{c}), and the number of proton transfer events per hydrogen atom and per picosecond (\textbf{d}) are shown for each oxoanion center atom type.
    Note that proton jumps were evaluated only every $100$th frame to reduce storage, so these values differ from those computed from the trajectories generated in the second stage of this investigation.
    }
    \label{fig:fig2}
\end{figure*}

The elemental composition of the remaining candidate materials is summarized in Fig.~\ref{fig:fig2}a.
The compositions are, unsurprisingly, dominated by oxygen and hydrogen atoms, alongside the predefined oxoanion center atoms P, S, Se, and As.
Beyond these elements, the dataset contains a wide range of alkali and alkaline-earth metals from groups $1$ and $2$ of the periodic table.
Several materials also include transition metals from groups $3$ to $12$ of the periodic table, most notably Zn, V, Mn, Ag, Pt, and Cd, as well as elements from groups $13$ to $18$ of the periodic table, including Al, C, Te, and Sn.
This compositional diversity demonstrates that restricting only the oxoanionic substructure enables exploration of an unusually broad chemical space for solid acids.

The distributions of key transport descriptors (see Methods) from the $100$~ps \textsc{MatterSim} simulations are shown in Fig.~\ref{fig:fig2}b--d.
While many materials exhibit very small proton diffusion coefficients $D_\mathrm{H}$ (cf.~Fig.~\ref{fig:fig2}b), a notable subset of S-, P-, and Se-based materials displays exceptionally high values exceeding $0.003$~{\AA}$^2\,\mathrm{ps}^{-1}$; for reference, the well-known solid acid CsH$_2$PO$_4$ exhibits $D_\mathrm{H}\sim0.001$~{\AA}$^2\,\mathrm{ps}^{-1}$ in its superprotonic regime \cite{struct_cdp, grunert2025}.

As shown in Fig.~\ref{fig:fig2}c, for each oxoanion center atom type, a large fraction of materials exhibit rapid anion rotation (i.e., $C_{\mathrm{XO}}(\tau=100~\mathrm{ps}) < 0.1$), although P-based systems show a comparatively higher fraction of slowly rotating anions.
The local proton transfer rate clusters tightly around $2.5$--$3.0$ events per H atom per picosecond (with few outliers) across all center-atom types (Fig.~\ref{fig:fig2}d).

Based on these results, our structurally motivated, motif-based screening identifies a focused set of strong candidates without relying on a large number of empirical descriptors.
From this initial screening, we selected the $70$ materials with the highest proton diffusion coefficients for further investigation.

\subsection*{Ab initio validation}

To obtain quantitatively accurate and fully converged diffusion coefficients, we performed $40$~ps AIMD simulations (see Methods) for each of the $70$ top candidates, generating system-specific training data.
These reference trajectories were used to fine-tune the \textsc{MACE-MP-0} foundation models for each material separately, enabling $3$~ns MD simulations at multiple temperatures ($400$~K, $500$~K, and $600$~K) while simultaneously retaining ab initio accuracy at multiple orders of magnitude lower computational cost \cite{mace_1, mace_2, mace_mp, hanseroth2025optimizing}.

\begin{figure}[ht]
    \centering
    \includegraphics{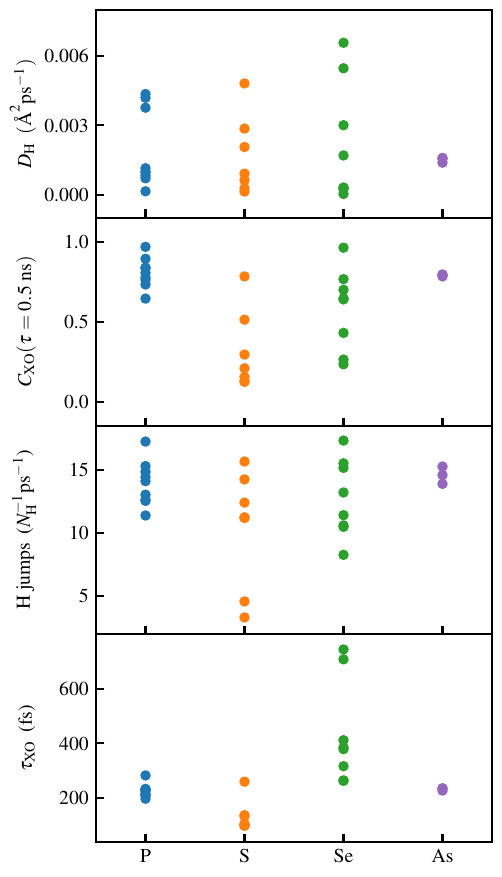}
    \caption{
    \textbf{Proton transport analysis based on long-time MD simulations using fine-tuned MLIPs.}
    From top to bottom: proton diffusion coefficients, oxoanion rotation dynamics, frequency of proton transfer events per hydrogen atom and per picosecond, and XO$_\mathrm{y}$ oscillation periods.
    Proton diffusion, oxoanion rotation, and proton-transfer frequencies are obtained from $400$~K $3$~ns MD trajectories generated with MLIPs fine-tuned to $40$~ps AIMD data for each material.
    XO$_\mathrm{y}$ oscillation periods are calculated from the $40$~ps AIMD trajectories, see Eq.~(\ref{eq:xoy_acf}) below.
    }
    \label{fig:fig3}
\end{figure}

The results reported below are based exclusively on the materials evaluated in this work and do not include any additional solid acid compounds.
The proton diffusion coefficients obtained from these long MD simulations are shown in Fig.~\ref{fig:fig3}.
The distribution of the proton diffusion coefficient $D_\mathrm{H}$ shows two distinct regimes: most materials cluster below $0.002$~{\AA}$^2\,\mathrm{ps}^{-1}$, while a subset exceeds $0.003$~{\AA}$^2\,\mathrm{ps}^{-1}$.
Critically, anion rotational dynamics, characterized by the  X--O bond-vector autocorrelation $C_\mathrm{XO}$, shown in Fig.~\ref{fig:fig3}, vary systematically with oxoanion identity:
Sulfur-centered oxoanions exhibit particularly fast rotation, while phosphorus- and arsenic-based materials display slower rotational dynamics, whereas selenium-based materials show a broad distribution.
The number of proton transfer events per hydrogen atom and per picosecond is approximately $14$ for materials with P-, \mbox{Se-,} and As-centered oxoanions, while sulfur-based materials show on average slightly lower values of $10$.
The oscillation periods of X--O bonds within the oxoanions are shown in Fig.~\ref{fig:fig3}.
Again, sulfur-based oxoanions exhibit the fastest dynamics, characterized by the shortest oscillation periods.
Phosphorus- and arsenic-based anions show intermediate behavior, whereas the slowest oscillations are observed for selenium-containing oxoanions.

\subsection*{Promising candidates}

From the $70$ materials identified in our two-stage screening and refinement workflow, we summarize the $27$ most promising candidate materials---defined as those with the highest proton diffusion coefficients---in Tab.~\ref{tab:tab1}.
For the full dataset of all $70$ materials, we refer to the Data Availability statement.
Most candidates exhibit activation energies of approximately $0.3\pm0.1$~eV, consistent with values reported for high-performance solid acid proton conductors (see Methods and Supplementary Note 6) \cite{struct_cdp, struct_cpp}.
The set includes both well-known solid acids and experimentally established proton conductors, as well as several previously unexplored materials.
Several of the high-performing candidates identified by the workflow correspond to well-established solid acids, which provides an important internal validation of the screening strategy.
LiH$_{3}$(SeO$_{3}$)$_{2}$ is a prominent example, as its proton motion and hydrogen-bond network were already investigated in the early 1970s using conductivity measurements, Raman spectroscopy, and neutron diffraction \cite{ultraslow_hydrogen_motion, al3230254_struc, al3230254_vib}.
These studies reported proton dynamics which are consistent with the high diffusion coefficients obtained here.
Similarly, CsHSeO$_{3}$ has been previously identified as a proton conductor based on infrared and Raman spectroscopic investigations, which indicated high proton mobility within its hydrogen-bonded framework \cite{al3230870_cond, al3230870_cond2}.
The reproduction of these known behaviors demonstrates that the applied motif-based screening and MD analysis reliably capture established Grotthuss-like proton transport mechanisms.

\begin{table}[ht]
    \centering
    \caption{
    \textbf{Candidate solid-acid proton conductors.}
    Summary of predicted proton transport descriptors for the shortlisted materials.
    Reported are the composition, the \textsc{Materials Project} (MP-ID) or \textsc{Alexandria} (Alex-ID) identifier, $D_\mathrm{H}$ (given in {\AA}$^2\,\mathrm{ps}^{-1}$), and literature references for the corresponding structures and/or conductivity measurements (Ref.).
    The proton diffusion coefficients $D_\mathrm{H}$ at $400$~K were obtained from $3$~ns MD trajectories at $400$~K computed using material-specific \textsc{MACE-MP-0} MLIPs fine-tuned to $40$~ps AIMD reference data (see Methods). 
    Multiple entries for the same composition represent different polymorphs.
    }
    \vspace{3mm}
    \begin{tabular}{lc@{\hspace{1em}}c@{\hspace{1em}}c@{\hspace{1em}}c@{\hspace{1em}}c@{\hspace{1em}}}
        \toprule
        Composition & MP-ID & $D_\mathrm{H}$ $(\text{\AA}^2 \mathrm{ps}^{-1})$  & Ref. \\
        \midrule
        KHSO$_{4}$ & 1199797 & 0.0048 & \cite{mp1199797_cond, mp1199797_cond2} \\
        (CH$_{3}$)$_{2}$NH$_{2}$H$_{2}$PO$_{4}$ & 1200282 & 0.0042 & \cite{mp1200282_struc, mp1200282_cond}  \\
        (NH$_{4}$)$_{3}$(HSeO$_{4}$)$_{2}$ & 1212963 & 0.0030 & --- \\
        K$_{4}$LiH$_{3}$(SO$_{4}$)$_{4}$ & 1196827 & 0.0029 & \cite{mp1196827_cond, mp1196827_cond2} \\
        H$_{56}$C$_{19}$S$_3$N$_8$O$_{15}$ & 1196861 & 0.0021 & \cite{mp1196861_struc} \\
        KH$_{2}$PO$_{4}$ & 699437 & 0.0008 & \cite{al3240111_struc} \\
        (Al(H$_2$O)$_6$)$_4$H$_4$(SO$_4$)$_7$ & 1229278 & 0.0006 & --- \\
        Tl$_2$H$_6$(SeO$_4$)(TeO$_6$) & 1196566 & 0.0002 & \cite{mp1196566_struc_cond} \\
        Rb$_2$H$_6$(SO$_4$)(TeO$_6$) & 559096 & 0.0001 & \cite{mp559096_struc_cond} \\
        \midrule
        Composition & Alex-ID & $D_\mathrm{H}$ $(\text{\AA}^2 \mathrm{ps}^{-1})$ & Ref. \\
        \midrule
        LiH$_{3}$(SeO$_{3}$)$_{2}$ & 3230254 & 0.0066 & \cite{al3230254_struc, al3230254_vib} \\ 
        CsHSeO$_{3}$ & 7647189 & 0.0055 & \cite{al3230870_cond, al3230870_cond2} \\
        Sn(H$_{2}$PO$_{4}$)$_{2}$ & 7541875 & 0.0044 & \cite{al7541875_struc} \\
        Ca(H$_{2}$PO$_{4}$)$_{2}$ & 5262379 & 0.0038 & \cite{al5262379_cond, al5262379_cond2} \\
        CsHSeO$_{3}$ & 3230870 & 0.0017 & \cite{al3230870_cond, al3230870_cond2}  \\
        K$_{2}$H$_{2}$As$_{2}$O$_{7}$ & 6308766 & 0.0016 & --- \\
        (KRb$_{3}$)(H$_{2}$As$_{2}$O$_{7}$)$_2$ & 6308815 & 0.0016 & --- \\
        (KCs$_{3}$)(H$_{2}$As$_{2}$O$_{7}$)$_2$ & 6308874 & 0.0014 & --- \\
        KH$_{2}$PO$_{4}$ & 3240111 & 0.0011 & \cite{al3240111_struc, al3240111_cond} \\
        Ba(H$_{2}$PO$_{4}$)$_{2}$ & 3239126 & 0.0010 & \cite{gilbert_struct, al3239126_cond} \\
        KH$_{2}$PO$_{4}$ & 3240113 & 0.0010 & \cite{al3240111_struc, al3240111_cond} \\
        CsHSO$_{4}$ & 3236698 & 0.0009 & \cite{al3236698_struc, al3236698_cond} \\
        (KCs$_{3}$)(H$_{2}$P$_{2}$O$_{7}$)$_2$ & 6308862 & 0.0007 & --- \\
        Rb$_{4}$H$_{4}$(Se$_{3}$S)O$_{16}$ & 6308968 & 0.0003 & --- \\
        RbHSeO$_{4}$ & 6308988 & 0.0003 & \cite{al6308988_struc, al6308988_cond} \\
        RbHSO$_{4}$ & 5291171 & 0.0003  & \cite{al5291171_struc, al5291171_cond} \\
        BaHPO$_{4}$ & 3280448 & 0.0001  & \cite{al3280448_struc, al3280448_cond} \\
        KH$_{3}$(SeO$_{3}$)$_{2}$ & 7754128 & 0.0001 & \cite{al7754128_struc_cond} \\
    \bottomrule
    \end{tabular}
    \label{tab:tab1}
\end{table}

Several industrially relevant and sustainable compounds also emerge among the top candidates.
Potassium bisulfate KHSO$_4$ is commercially produced and widely used, for example as a precursor in wine-making, and its proton conductivity has been experimentally characterized for decades \cite{mp1199797_cond, mp1199797_cond2}.
Calcium dihydrogenphosphate Ca(H$_{2}$PO$_{4}$)$_{2}$ is another well-known compound that is industrially relevant as a fertilizer and was experimentally shown to exhibit high proton conductivity already in the late 1980s \cite{al5262379_cond, al5262379_cond2}.
KH$_{2}$PO$_{4}$ represents a canonical solid acid with well-documented super-protonic behavior, which has been extensively investigated using techniques such as proton NMR \cite{al3240111_struc, al3240111_cond, ultraslow_hydrogen_motion}.
Barium dihydrogenphosphate Ba(H$_{2}$PO$_{4}$)$_{2}$ also belongs to this class of experimentally validated super-protonic conductors and was also rediscovered by {\v{Z}}guns et al.~\cite{vzguns2024uncovering}, confirming its exceptional transport descriptors \cite{gilbert_struct, al3239126_cond}.

Beyond known systems, our screening found several materials that have been synthesized but whose proton-transport descriptors have not yet been characterized to the best of our knowledge, as they have not yet been investigated or their properties could only be derived indirectly.
Sn(H$_{2}$PO$_{4}$)$_{2}$ is one such case: its crystal structure has been reported, yet direct conductivity measurements are, to the best of our knowledge, unavailable \cite{al7541875_struc}.
Nevertheless, it is closely related to SnHPO$_4$ which exhibits high proton diffusion \cite{al7541875_cond}.
A similar situation is found for (NH$_{4}$)$_{3}$(HSeO$_{4}$)$_{2}$, which has not been experimentally characterized with respect to proton conductivity, despite the fact that many of its structural derivatives are known to exhibit high proton mobility \cite{mp1212963_deriv, mp1212963_deriv2, mp1212963_deriv3}.
These compounds therefore represent particularly attractive targets for experimental validation because they offer a more sustainable materials basis---being composed mainly of light, earth-abundant elements (e.g., C, H, O, and N) rather than relying on heavy/less abundant alkali metals such as Cs or Rb---and because their proton-transport behavior remains largely unexplored.

Organic and hybrid materials also appear as promising candidates.
(CH$_{3}$)$_{2}$NH$_{2}$H$_{2}$PO$_{4}$ is an organic crystal that undergoes a phase transition at $259$~K and exhibits high proton conductivity in experiment \cite{mp1200282_struc, mp1200282_cond}.
Its successful identification highlights that the applied motifs are sufficiently general to capture proton-conducting mechanisms beyond purely inorganic frameworks.
Tetra\-ethyl\-ammonium hydrogen sulfate urea H$_{56}$C$_{19}$S$_3$N$_8$O$_{15}$ has been structurally characterized but its proton conductivity has not yet been experimentally investigated \cite{mp1196861_struc}.
The present results suggest that such complex hydrogen-bonded organic salts may represent an underexplored class of solid acid-like proton conductors.

A number of arsenate- and tellurium-containing compounds also display favorable proton transport properties, although their chemical composition raises sustainability and toxicity concerns.
K$_{2}$H$_{2}$As$_{2}$O$_{7}$ and its structurally related mixed-cation derivatives (KRb$_{3}$)(H$_{2}$As$_{2}$O$_{7}$)$_2$, (KCs$_{3}$)(H$_{2}$As$_{2}$O$_{7}$)$_2$, and (KCs$_{3}$)(H$_{2}$P$_{2}$O$_{7}$)$_2$ have not been evaluated experimentally with respect to proton conductivity.
Their appearance among the high-performing candidates indicates that arsenate-based frameworks can support efficient proton transport, albeit at the cost of reduced environmental compatibility.
Tellurium-containing compounds such as Rb$_2$H$_6$(SO$_4$)(TeO$_6$) and Tl$_2$H$_6$(SeO$_4$)(TeO$_6$) exhibit super-protonic phases similar to the well-known solid acid Cs$_2$H$_6$(SO$_4$)(TeO$_6$) \cite{mp559096_struc_cond, mp559096_deriv, mp1196566_struc_cond}.

Among lower-performing materials, several known solid acids are recovered, further supporting the robustness of the workflow.
CsHSO$_{4}$, a prominent derivative of CsH$_2$PO$_4$, is a well-established solid acid with experimentally confirmed proton conductivity \cite{al3236698_struc, al3236698_cond}.
RbHSeO$_{4}$ and RbHSO$_{4}$ are also well-known solid acids with documented transport properties \cite{al6308988_struc, al6308988_cond, al5291171_struc, al5291171_cond}.
Rb$_4$H$_4$(Se$_3$S)O$_{16}$ has not been reported previously, but its close structural similarity to the mixed sulfate-selenate conductor RbH(SO$_4$)$_{0.81}$(SeO$_4$)$_{0.19}$ suggests that related proton transport mechanisms may be active \cite{al6308968_deriv}.
BaHPO$_{4}$ is another known proton conductor that is used as a partial substitution in CsH$_2$PO$_4$ to tune its transport properties \cite{al3280448_struc, al3280448_cond, al3280448_deriv}.
KH$_3$(SeO$_3$)$_2$ completes the set of filtered systems and is a well-established solid acid with experimentally observed proton dynamics \cite{al7754128_struc_cond, ultraslow_hydrogen_motion}.

Taken together, the discussed materials include well-established solid acids as well as less-explored candidates.
Due to the high recovery rate of known proton conductors, investigating the previously uncharacterized compounds in more depth, both theoretically and experimentally, should be a worthwhile endeavor.

\subsection*{Structural origin of proton transfer}

\begin{figure*}[ht]
    \centering
    \includegraphics{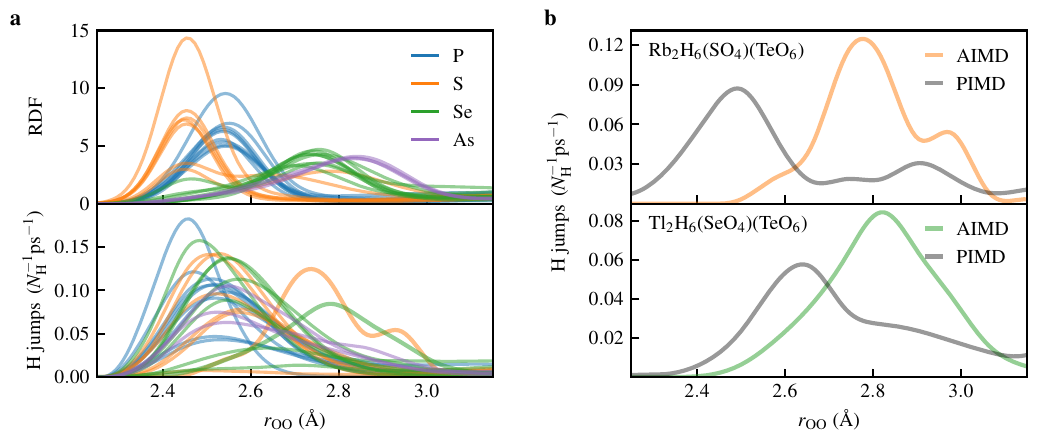}
    \caption{\textbf{Structural origin of proton transfer.} 
    \textbf{a} Oxygen--oxygen radial distribution functions (RDF, top) and distribution of proton transfer events as a function of instantaneous oxygen--oxygen distance (bottom) obtained from the $40$~ps AIMD trajectories. 
    \textbf{b} Comparison of the proton transfer distance distributions obtained from AIMD and PIMD simulations for the outlier materials containing the (TeO$_6$)-group: Rb$_2$H$_6$(SO$_4$)(TeO$_6$) (top) and Tl$_2$H$_6$(SeO$_4$)(TeO$_6$) (bottom).}
    \label{fig:fig4}
\end{figure*}

With the large, internally consistent set of solid-acid trajectories generated through our workflow---spanning many materials that are already known and synthetically accessible---we can go beyond ranking candidates and instead ask what generic structural features control proton transfer in solid acids. 
Motivated by this, we reanalyzed the AIMD trajectories to connect local geometry to hopping activity, focusing in particular on the instantaneous oxygen--oxygen separation.

In the top panel of Fig.~\ref{fig:fig4}a, we show the oxygen--oxygen radial distribution function (RDF) for all $27$ candidate solid acid proton conductors found through our two-stage high-throughput workflow.
Materials sharing the same anion center (P, S, Se, As) exhibit very similar oxygen--oxygen RDFs, reflecting that the local oxoanion geometry is primarily determined by the identity of the central atom.
The RDF maxima occur at O-O distances of approximately $2.45$,  $2.55$,  $2.75$, and $2.85$~{\AA} for S, P, Se, and As oxoanions, respectively.

The bottom panel of Fig.~\ref{fig:fig4}a presents the normalized distributions of proton transfer events as a function of the instantaneous oxygen--oxygen distance.
Remarkably, across all compositions and oxoanion center atom types, proton transfer occurs preferentially at a characteristic oxygen--oxygen separation of approximately $2.5$~{\AA}, regardless of the local oxygen--oxygen distribution inherent to each material.
Even in selenium- and arsenic-based materials where the oxygen RDFs show only weak intensity near this distance, transfer events cluster at $2.5$~{\AA}.

This finding suggests that proton transfer occurs preferentially at an oxygen--oxygen separation of approximately $2.5$~{\AA} across the diverse solid-acid chemistries investigated here, providing the mechanistic insight that this separation represents a largely structure-independent geometric condition for efficient hopping.
Only two outlier materials (Rb$_2$H$_6$(SO$_4$)(TeO$_6$) and Tl$_2$H$_6$(SeO$_4$)(TeO$_6$)) bearing tellurium-based substructures (TeO$_6$ groups) display shifted transfer distances, reflecting the altered bonding environment around large tellurium centers.

Strikingly, when nuclear quantum effects are included via path-integral molecular dynamics (PIMD), even these outliers restore the characteristic $2.5$~{\AA} transfer distance, as shown in Fig.~\ref{fig:fig4}b.
The inclusion of nuclear quantum effects leads to a redistribution of proton probability density toward shorter oxygen--oxygen separations, recovering the universal transfer distance and suggesting that the $2.5$~{\AA} motif represents a fundamental length scale in proton-conducting solids.

\section*{Discussion}

In this study, we combined motif-based structural screening with machine-learning-driven MD simulations to identify solid-acid proton conductors across a large chemical space.
Starting from over six million crystal structures, we applied two complementary structural motifs derived from known solid acids---hydrogen coordination by multiple oxoanion oxygens and oxygen atoms with multiple hydrogen neighbors---to filter candidates for materials with solid-state proton transport.
Subsequent MD simulations with the \textsc{MatterSim} foundation model, followed by fine-tuned \textsc{MACE-MP-0} models trained on AIMD reference data, enabled quantitative ranking of proton diffusion coefficients and mechanistic insights into the underlying transport dynamics.
The workflow successfully recovered well-established solid acids such as CsHSO$_4$, KHSO$_4$, and KH$_2$PO$_4$ while simultaneously identifying previously unexplored candidates including organic salts, ammonium-based selenates, and tin phosphates.
Among the $27$ highest-performing materials, several exhibit diffusion coefficients exceeding $0.003$~{\AA}$^2\,\mathrm{ps}^{-1}$ at $400$~K, comparable to or surpassing well-known superprotonic conductors.

The comparative analysis of the proton transport descriptors across all candidates reveals that efficient proton conduction requires more than favorable structural motifs alone.
While the local proton transfer frequency remains nearly constant across all oxoanion center-atom types (S, P, Se and As), macroscopic diffusion coefficients vary by more than an order of magnitude.
This decoupling demonstrates that long-range proton transport is governed primarily by the connectivity and persistence of hydrogen-bond networks rather than by the intrinsic hopping rate at individual sites.
Anion rotational dynamics emerge as a critical mediator: selenium-based materials, which exhibit intermediate rotation rates and oscillation periods, achieve the highest average diffusion coefficients, suggesting an optimal balance between hydrogen-bond breaking and reformation.
In contrast, sulfur-centered oxoanions rotate very rapidly but show fewer transfer events, potentially disrupting network connectivity, while phosphorus- and arsenic-based systems display slower reorientation that may limit pathway regeneration.

AIMD simulations show that proton transfer between neighboring oxoanions occurs at a well-defined oxygen--oxygen separation of about $2.5$~{\AA}.
This characteristic distance is consistently observed across all investigated crystal structures, irrespective of the oxoanion center atom type and cationic species.
Even in Se- and As-based materials, where the oxygen--oxygen RDFs exhibit only weak intensity near this distance, proton transfer events occur preferentially at oxygen--oxygen distances of about $2.5$~{\AA}.
Notably, this implies that fast proton transfer is controlled by whether the structure can transiently realize a specific hydrogen-bond geometry---an instantaneous oxygen--oxygen separation of approximately $2.5$~{\AA} at the moment of transfer---rather than by the average (most probable) oxygen--oxygen distance set by the host crystal lattice.
In other words, even if a material's oxygen--oxygen RDF peaks at larger separations, thermal fluctuations and local distortions can occasionally compress donor and acceptor oxygens into this ``transfer-ready'' configuration, and proton hops occur predominantly during these geometrically optimal events.
This supports the concept of a structure-independent optimal proton transfer distance: different oxoanion chemistries may differ in how often they sample oxygen--oxygen distance of approximately $2.5$~{\AA}, but the transfer geometry itself remains universal for solid-acids.
To the best of our knowledge, this is the first systematic demonstration of such a universal transfer distance across a diverse set of solid-acid chemistries and crystal structures.
Two outlier compounds containing Te-based substructures (TeO$_6$) display slightly shifted transfer distances in classical MD simulations, reflecting the altered bonding environment around the large tellurium centers.
When nuclear quantum effects are included through PIMD simulations, the universal transfer distance of approximately $2.5$~{\AA} is restored (see Supplementary Note 7).

The identified candidate list includes well-known solid acids such as KHSO$_4$, KH$_2$PO$_4$, Ca(H$_2$PO$_4$)$_2$, and CsHSO$_4$, validating the workflow, as well as previously uncharacterized compounds such as Sn(H$_2$PO$_4$)$_2$, (NH$_4$)$_3$(HSeO$_4$)$_2$, K$_2$H$_2$As$_2$O$_7$, and complex organic salts (e.g., H$_{56}$C$_{19}$S$_3$N$_8$O$_{15}$) that emerge as promising compounds.
The successful recovery of experimentally established conductors demonstrates that structural motifs combined with dynamical screening can reliably identify Grotthuss-like transport mechanisms without requiring extensive empirical descriptors or prior knowledge of specific compositions.
At the same time, the presence of multiple uncharacterized yet structurally validated compounds highlights opportunities for targeted experimental follow-up to expand the solid-acid design space beyond conventional phosphate and sulfate chemistries.

Overall, this work demonstrates that motif-based structural filtering combined with machine-learning-driven MD simulations enables the screening of proton-conducting solids beyond conventional chemical design rules while providing mechanistic insight into the interplay between anion dynamics and proton transport---an approach that is now feasible, mainly thanks to the advent of MLIP foundation models and their efficient fine-tuning using modest AIMD reference datasets.

\section*{Methods}

In this study, the simulations required roughly $2.5$ million CPU hours and $30.1$ thousand GPU hours, yielding a total of approximately $0.7$ microseconds of ab initio-accuracy MD trajectories.
The high-throughput MLIP simulation workflow (including input generation and post-processing) was automated using our toolkit \cite{haenseroth2025amaceingtoolkit} (see Code Availability statement).
For the data supporting this work, please refer to the Data Availability statement.

\subsection*{High-throughput \textsc{MatterSim} MD simulations}

The initial screening of materials via MD simulations was performed using \textsc{MatterSim} (version 1.1.2) with the \textsc{MatterSim}-v1.0.0-5M foundation model \cite{mattersim}.
\textsc{MatterSim} is an invariant graph neural network based on the \textsc{M3GNet} architecture \cite{mattersim, m3gnet}.
The foundation model \textsc{MatterSim}-v1.0.0-5M was trained by the \textsc{Microsoft Research AI for Science Team} using a DFT dataset not publicly available, incorporating approximately $17$~million materials across a temperature range of $0$~K to 5,000 K and a pressure range of $0$~GPa to 1,000 GPa \cite{mattersim}.
MD simulations of 3,967 materials were conducted on CPU at $400$~K for $100$~ps with a timestep of $0.5$~fs.
Every $100$th trajectory frame was used for the structural and dynamical analyses described below.

\subsection*{AIMD simulations}

Reference AIMD simulations were performed on CPUs using CP2K (version 2025.1) with the PBE exchange-correlation functional, GTH pseudopotentials, DZVP-MOLOPT basis set, and the Nos\'{e}-Hoover chain thermostats \cite{cp2k_1, cp2k_2, cp2k_3, cp2k_4, cp2k_5, cp2k_quickstep, cp2k_basis-set, cp2k_orb_trans, cp2k_gth-pseudopot1, cp2k_gth-pseudopot2, cp2k_gth-pseudopot3, pbe, nose1, nose2, nose3}.
The AIMD simulations were carried out for $40$~ps at $400$~K with a timestep of $0.5$~fs for $70$ materials.
However, for a subset of materials the resulting AIMD trajectories showed decomposition and/or the formation of water or hydrogen (see Supplementary Note 3), or the simulations failed to converge.
As a result, $27$ materials with stable and converged AIMD trajectories were retained for subsequent analysis.

\subsection*{\textsc{MACE} fine-tuning and MD simulations}

The fine-tuning of the foundation model, as well as the subsequent MD simulations employing the fine-tuned potential, were performed using the \textsc{MACE-torch} Python package (version 0.3.14) \cite{mace_1,mace_2,mace_mp}. 
Training data consisted of $500$ configurations per system extracted from first 60,000 frames of the AIMD trajectories by equidistant sampling.
For each material, the foundation model \textsc{MACE}-MP-0-small was fine-tuned on the material-specific ab initio datasets generated through the AIMD simulation described above.
This procedure yielded $27$ fine-tuned MACE models.
The model evaluation was done on $200$ configurations from the last 20,000 frames of the AIMD trajectories. 
The force errors of the resulting models did not exceed $70.9$~meV/{\AA} (mean $\pm$ standard deviation: $44.2\pm10.5$~meV/{\AA}), and the energy errors remained below $3.4$~meV/atom (mean $\pm$ standard deviation: $1.5\pm0.9$~meV/atom). 

MD simulations were performed with \textsc{ASE} using Nos\'{e}-Hoover chain thermostats set up with the \textsc{aMACEing\_toolkit} \cite{mace_1, mace_2, ase, nose1, nose2, nose3, haenseroth2025amaceingtoolkit, ase} for $3$~ns at $400$~K, $500$~K, and $600$~K with a timestep of $0.5$~fs.
Fine-tuning and MD simulations were performed using \textsc{NVIDIA} A100 GPUs.

\subsection*{\textsc{MACE} PIMD simulations}

PIMD simulations were performed using \textsc{i-PI} interfaced with \textsc{LAMMPS} and the fine-tuned \textsc{MACE} models \cite{i-PI1_0, i-PI2_0, i-PI3_0}.
The simulations employed $8$ beads to represent the quantum mechanical nature of protons, with a simulation duration of $10$~ps and a timestep of $0.5$~fs.
The PIGLET thermostat was employed with parameters optimized for $8$ beads and a target temperature of $400$~K \cite{PIGLET1, PIGLET2}.
Thermostat parameters were generated with the \textsc{GLE4MD} tool with spectral range extending to $\omega_\mathrm{max}=4000~\mathrm{cm}^{-1}$ and $\omega_\mathrm{max}/\omega_\mathrm{min}=10^4~\mathrm{cm}^{-1}$ \cite{gle4md}.
The PIMD simulations were performed using \textsc{NVIDIA} A100 GPUs.

\subsection*{Trajectory analyses}

To obtain converged diffusion coefficients, fine-tuned foundation models were employed, following previously established protocols for the solid acids CsH$_2$PO$_4$ and Cs$_7$(H$_4$PO$_4$)(H$_2$PO$_4$)$_8$ \cite{grunert2025}:
We calculated the mean square displacement (MSD) of the protons in both \textsc{MatterSim} and fine-tuned \textsc{MACE} foundation model trajectories according to
\begin{equation}
    \mathrm{MSD}(\tau) = \big\langle \left\vert \mathbf{r}_{i}(t+\tau) - \mathbf{r}_{i}(t) \right\vert^{2} \big\rangle_{t,i}
    \label{eq:msd}
\end{equation}
Here, $\mathbf{r}_{i}(t)$ denotes the position of particle $i$ at time $t$. 
The $\mathrm{MSD}(\tau)$ was computed by averaging over all protons $i$ and all initial times $t$ satisfying $t + \tau < t_\mathrm{MD}$, where $t_\mathrm{MD}$ is the total simulation time.
Then, diffusion coefficients were derived by linear fitting
\begin{equation}
    D = \frac{1}{6} \frac{d}{d \tau}\mathrm{MSD}(\tau)
    \label{eq:diff_coeff}
\end{equation}
within the time interval $\tau \in [10~\mathrm{ps},\,30~\mathrm{ps}]$.

Proton transport proceeds through transfer events between two XO$_{\mathrm{y}}$ groups, with $\mathrm{X}\in\{\mathrm{P},\,\mathrm{S},\,\mathrm{Se},\,\mathrm{As}\}$ and $\mathrm{y}\in\{3,4\}$, where a proton covalently bound to one XO$_{\mathrm{y}}$ and hydrogen-bonded to the other exchanges these bonding roles, thereby hopping between the groups.
Therefore, the frequency of proton transfer events per hydrogen atom and per picosecond was evaluated.
The second part of the Grotthuss mechanism in the solid acids involves reorientation of the XO$_{\mathrm{y}}$ groups, facilitating subsequent proton transfer events with different XO$_{\mathrm{y}}$ group partners. 
The rotational dynamics of the XO$_{\mathrm{y}}$ groups were analyzed using autocorrelation function of the unit vector $\mathbf{r}_{\mathrm{XO}}$  along the X--O bond:
\begin{equation}
    C_{\mathrm{XO}}(\tau) = \big\langle \langle \mathbf{r}_{\mathrm{XO}}(t_0) \cdot \mathbf{r}_{\mathrm{XO}}(t_0 + \tau) \rangle_{\mathrm{XO}} \big\rangle_{t_0}
    \label{eq:vacf}
\end{equation}

The X--O$_{\mathrm{y}}$ oscillation period $\tau_{XO}$ was determined from the dominant vibrational mode of the average X--O bond distance within each XO$_{\mathrm{y}}$ polyhedron as follows.
For each X center, the instantaneous X--O average distance was defined as
\begin{equation}
    d_\mathrm{X}(t) = \frac{1}{\bigl\vert N_\mathrm{O}(\mathrm{X};r_c)\bigr\vert} \sum_{j \in N_\mathrm{O}(\mathrm{X};r_c)} \bigl\vert \mathbf{r}_{\mathrm{XO}_j}(t) \bigr\vert
    \label{eq:xoy_avg_distance}
\end{equation}
where $N_\mathrm{O}(\mathrm{X};r_c)$ is the set of oxygen atoms within a cutoff radius $r_c$ bonded to X, and $\bigl\lvert \mathbf{r}_{\mathrm{XO}_j}(t) \bigr\rvert$ is the X--O$_j$ distance at time $t$.
After subtracting the time average $\bar{d}_\mathrm{X} = \langle d_\mathrm{X}(t)\rangle_t$, the normalized distance autocorrelation function was computed as
\begin{equation}
    C_{\mathrm{XO}_\mathrm{y}}(\tau) = \frac{\big\langle \delta d_\mathrm{X}(t) \delta d_\mathrm{X}(t+\tau)\big\rangle_{t,\mathrm{X}}}{\big\langle \delta d_\mathrm{X}(t)^2\big\rangle_{t,\mathrm{X}}}
    \label{eq:xoy_acf}
\end{equation}
with $\delta d_\mathrm{X}(t) = d_\mathrm{X}(t) - \bar{d}_\mathrm{X}$.
Here, $\langle\cdot\rangle_{t,\mathrm{X}}$ denotes averaging over time origins and all XO$_{\mathrm{y}}$ units.
The corresponding power spectrum was obtained via discrete Fourier transform as
\begin{equation}
    S_{\mathrm{XO}_\mathrm{y}}(f) = \left|\mathcal{F}\{C_{\mathrm{XO}_\mathrm{y}}(\tau)\}(f)\right|^2,
    \label{eq:xoy_spectrum}
\end{equation}
Finally, the X--O$_\mathrm{y}$ oscillation period $\tau_\mathrm{XO}$ was defined as $\tau_\mathrm{XO} = 1/f_\mathrm{max}$, where \(f_\mathrm{max}\) is the position of the dominant finite-frequency peak of the spectrum.

Activation energies $E_a$ for proton diffusion were obtained from the diffusion coefficients calculated from the $3$~ns fine-tuned MLIP MD trajectories at $400$~K, $500$~K, and $600$~K. 
The temperature dependence of $D$ is assumed to follow Arrhenius behavior
\begin{equation}
    D(T) = A \exp\left( -\frac{E_a}{k_b T} \right)
    \label{eq:act_ener}
\end{equation}
with a prefactor $A$, an activation energy $E_a$, the Boltzmann constant $k_b$, and the absolute temperature $T$.

\section*{Data Availability}

The dataset containing crystal structures, the fine-tuning data, and the resulting models are available at \url{doi.org/10.5281/zenodo.18315233}.

\section*{Code Availability}

The used third-party codes \textsc{CP2K}, \textsc{MACE}, \textsc{MatterSim} are available at \url{cp2k.org}, \url{github.com/acesuit/mace} and \url{github.com/microsoft/mattersim}, respectively. 
The high-throughput input creator used in this study is available at \url{github.com/jhaens/amaceing_toolkit}.

\section*{Acknowledgments}

We thank the staff of the Compute Center of the Technische Universität Ilmenau, especially Mr.~Henning~Schwanbeck for providing an excellent research environment. 
M.~Großmann thanks J.~Lobert for the helpful suggestions that improved the design of Fig.~1.
This work is supported by the Carl-Zeiss-Stiftung (SustEnMat, funding code: P2023-02-008), the Thüringer Aufbaubank (TAB) (KapMemLyse, grant no.~2024 FGR 0081 / 0082) and the European Social Fund Plus (ESF+). 

\section*{Competing interests}

The authors declare no competing interests.

\section*{Author contributions}
M.~Großmann conceived the idea; M.~Großmann filtered the \textsc{Materials Project} and the \textsc{Alexandria} database based on chemical motifs suggested by C.D., J.H.~wrote the high-throughput workflow, performed all calculations; J.H., C.D., M.G., and M.G.~analyzed the data; M.~Großmann visualized all results and J.H.~wrote the first draft of the manuscript. 
E.R.~and C.D.~supervised the work; all authors revised and approved the manuscript.

\bibliography{literature}

\end{document}


\title{Supplementary Information for "High-throughput screening and mechanistic insights into solid acid proton conductors"}

\author{Jonas H{\"a}nseroth}
\email{jonas.haenseroth@tu-ilmenau.de}
\affiliation{Institute of Physics and Institute of Micro- and Nanotechnologies, Technische Universit\"at Ilmenau, 98693 Ilmenau, Germany}

\author{Max Gro{\ss}mann}
\affiliation{Institute of Physics and Institute of Micro- and Nanotechnologies, Technische Universit\"at Ilmenau, 98693 Ilmenau, Germany}

\author{Malte Grunert}
\affiliation{Institute of Physics and Institute of Micro- and Nanotechnologies, Technische Universit\"at Ilmenau, 98693 Ilmenau, Germany}

\author{Erich Runge}
\affiliation{Institute of Physics and Institute of Micro- and Nanotechnologies, Technische Universit\"at Ilmenau, 98693 Ilmenau, Germany}

\author{Christian Dre{\ss}ler}
\affiliation{Institute of Physics and Institute of Micro- and Nanotechnologies, Technische Universit\"at Ilmenau, 98693 Ilmenau, Germany}

\date{\today}

\maketitle

For more information on abbreviations, please refer to the main text, where all abbreviations are defined in detail.
Abbreviations not introduced in the main text are defined here.

\tableofcontents
\newpage
\clearpage

\section*{Supplementary Note 1: Elemental composition of the candidate set}

\begin{figure}[ht]
    \centering
    \includegraphics{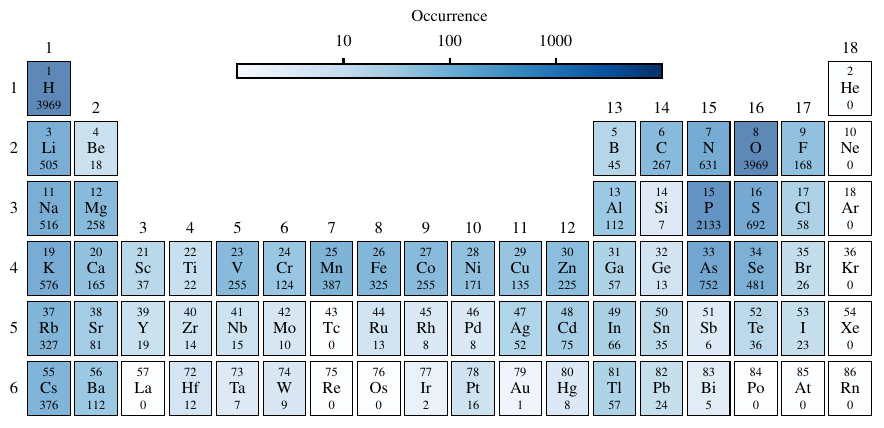}
    \caption{
    \textbf{Elemental diversity of the candidate set obtained through motif-based structural filtering.}
    Periodic-table heat map summarizing the elemental composition of the 3,969 candidates extracted by the motif search from the \textsc{Alexandria} \cite{alexandria, cavignac2025ai} and \textsc{Materials Project} \cite{mp_1, mp_2} databases.
    Colors encode how often each element appears among the filtered structures, with the exact counts printed below the corresponding element symbols.
    The lanthanides and the seventh period are omitted because no candidates in the motif-filtered set contain elements from these groups.
    }
    \label{si_fig:1}
\end{figure}

\newpage
\clearpage

\section*{Supplementary Note 2: Distribution of the simulation cell size of the candidate set}

\begin{figure}[ht]
    \centering
    \includegraphics{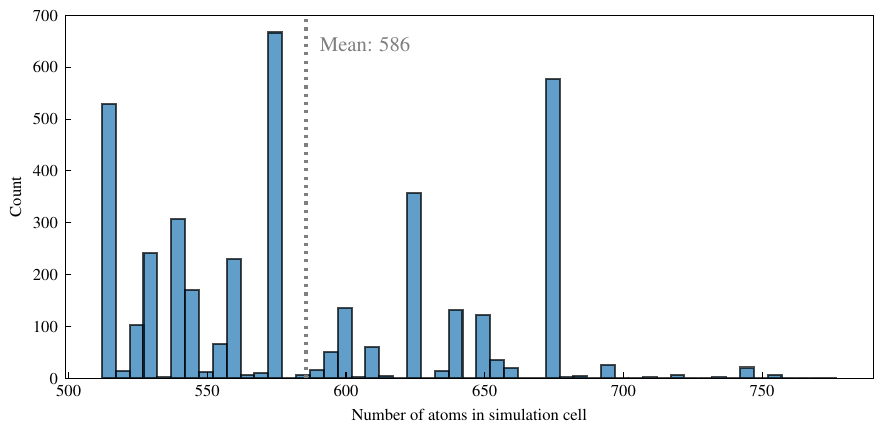}
    \caption{
    \textbf{Distribution of simulation cell sizes used for screening.}
    Histogram of the number of atoms in the simulation supercells constructed for the 3,969 candidates obtained through motif-based structural filtering of the \textsc{Materials Project} \cite{mp_1, mp_2} and \textsc{Alexandria} \cite{alexandria, cavignac2025ai} database.
    The cell-size distribution reflects the use of enlarged supercells to obtain statistically meaningful proton-transport and anion-dynamics descriptors from high-throughput MD simulations (see Methods in the main text).
    }
    \label{si_fig:2}
\end{figure}

\section*{Supplementary Note 3: Water and hydrogen molecule detection}

Molecule detection was performed using the \texttt{molecule\_recognition} function implemented in the \textsc{pbctools} package (\url{github.com/jhaens/pbctools}). 
Each hydrogen atom is assigned to its nearest neighbor, and bonds between atoms X and Y are identified using a distance cutoff based on element-specific van der Waals radi.
If H$_2$O or H$_2$ molecules were detected, the corresponding material was excluded from further analysis.

\newpage
\clearpage

\section*{Supplementary Note 4: Elemental composition of the fine-tuning candidate set}
\begin{figure}[ht]
    \centering
    \includegraphics{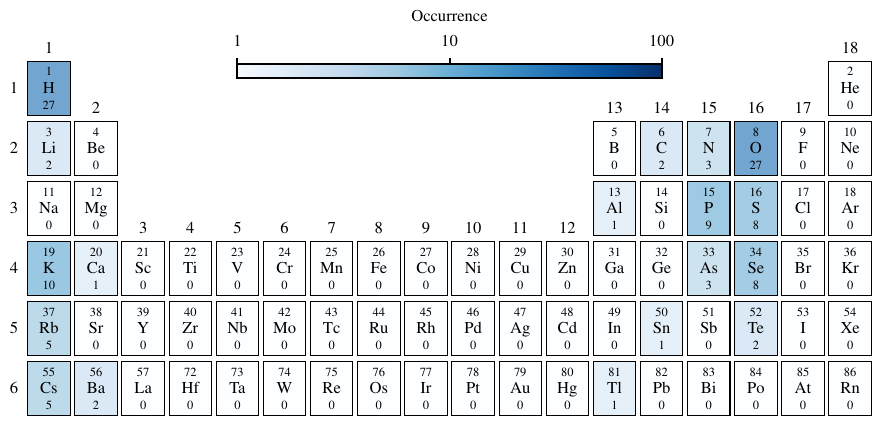}
    \caption{
    \textbf{Elemental composition of the fine-tuning candidate set.}
    Periodic-table heat map showing the elemental composition of the $27$ most promising candidate materials for which material-specific fine-tuned \textsc{MACE} models were trained \cite{mace_1, mace_2, mace_mp} (see Methods in the main text).
    Colors indicate the number of occurrences of each element within this final set, with exact counts reported below the corresponding symbols.
    The lanthanides and the seventh period are omitted because they do not occur in the fine-tuning set. 
    }
    \label{si_fig:3}
\end{figure}

\newpage
\clearpage

\section*{Supplementary Note 5: Simulation cells of the final candidates}

\begin{figure}[h!]
    \centering
    \includegraphics{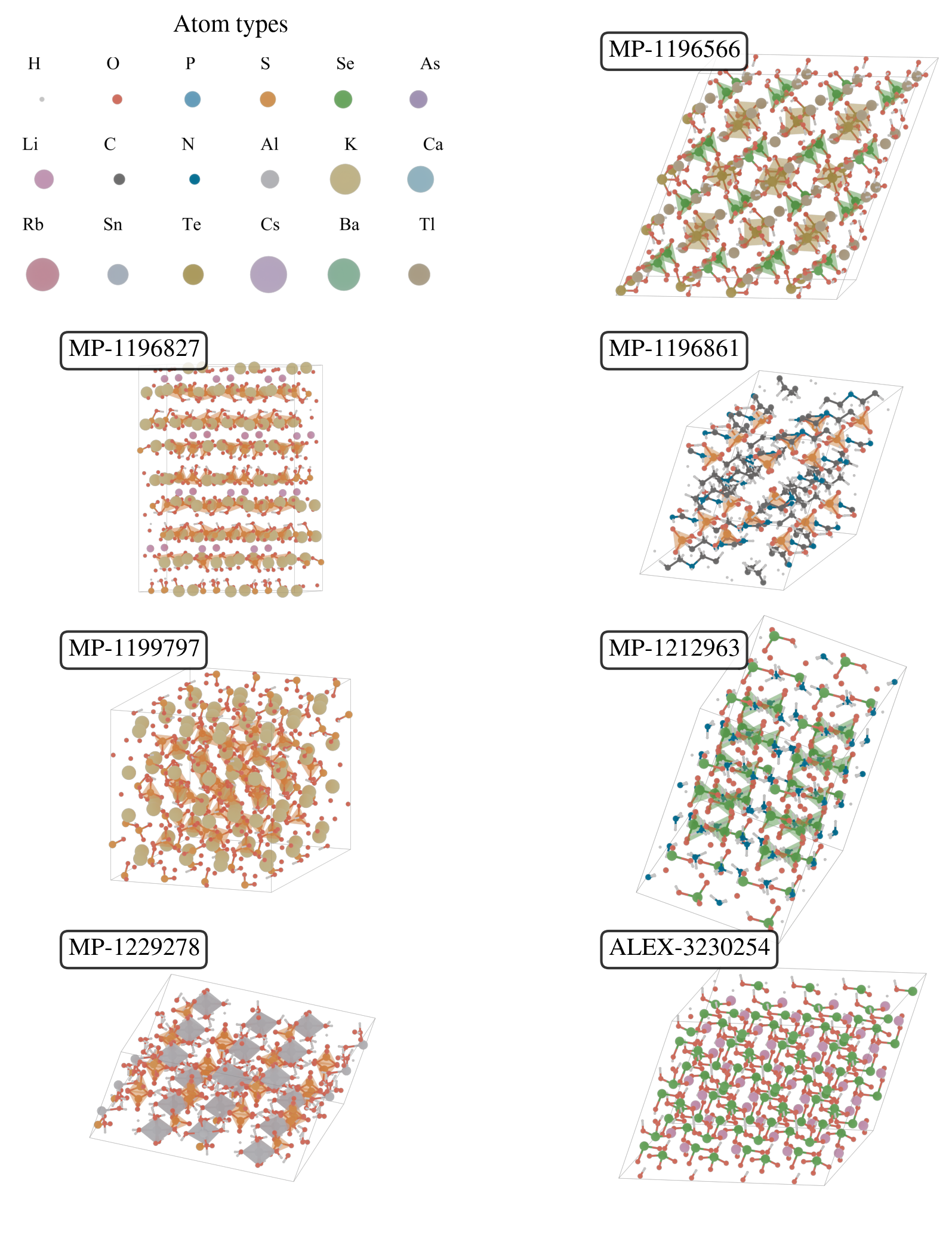}
    \caption{
    \textbf{Simulation cells of the final candidate materials.}
    Visualization of the simulation supercells used for the final simulations and  analyses.
    The upper-left panel provides the atom-type legend, and titles above each structure report the corresponding Materials Project- or Alexandria-identifier.
    Oxoanions are highlighted as coordination polyhedra, colored by the respective center atom.
    Periodic cell boundaries are indicated by black outlines.
    }
    \label{si_fig:4}
\end{figure}

\begin{figure}[ht]
    \centering
    \includegraphics{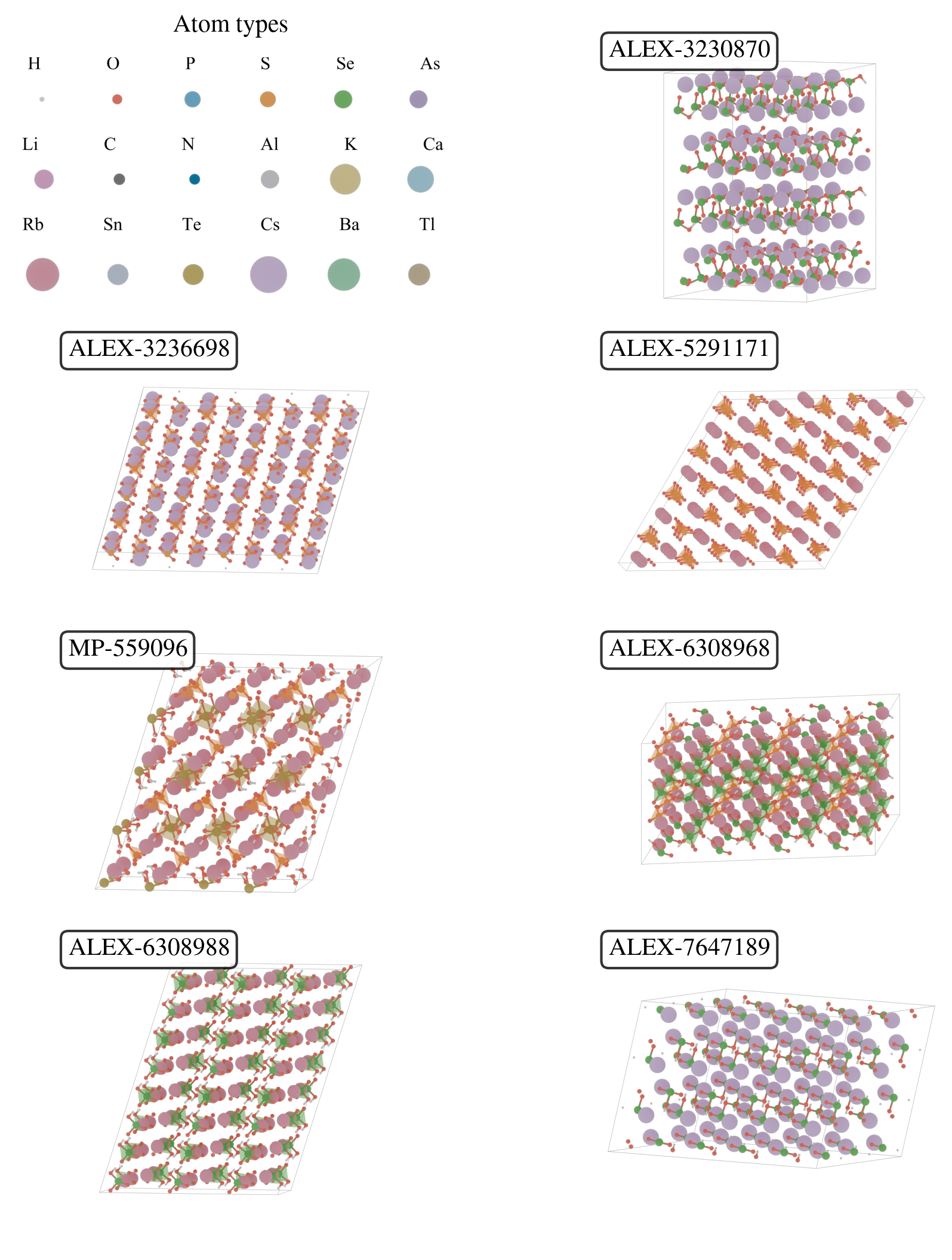}
    \caption{
    \textbf{Simulation cells of the final candidate materials.}
    Extension of Supplementary Fig.~\ref{si_fig:4}.
    }
    \label{si_fig:5}
\end{figure}

\begin{figure}[ht]
    \centering
    \includegraphics{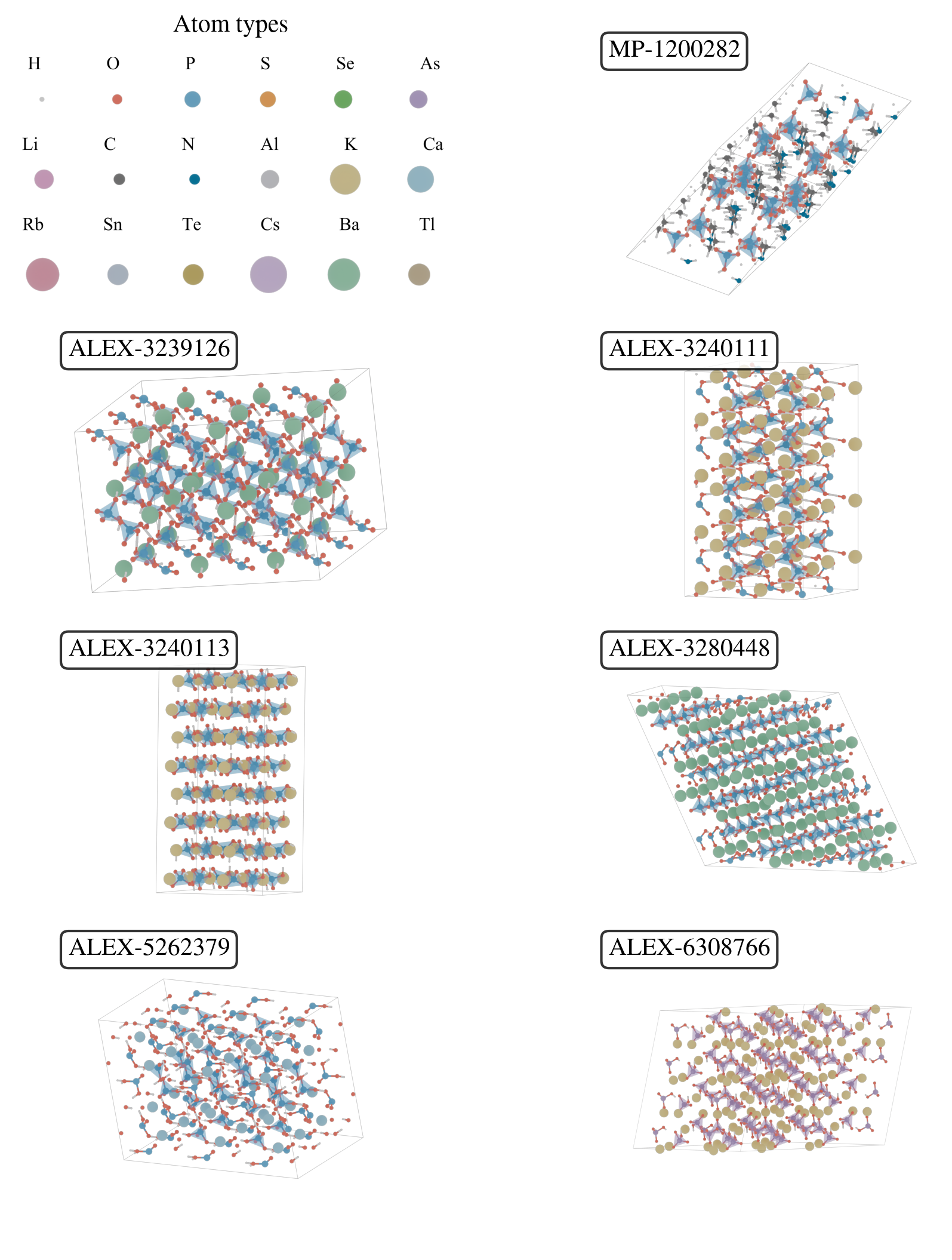}
    \caption{
    \textbf{Simulation cells of the final candidate materials.}
    Extension of Supplementary Fig.~\ref{si_fig:4}.
    }
    \label{si_fig:6}
\end{figure}

\begin{figure}[ht]
    \centering
    \includegraphics{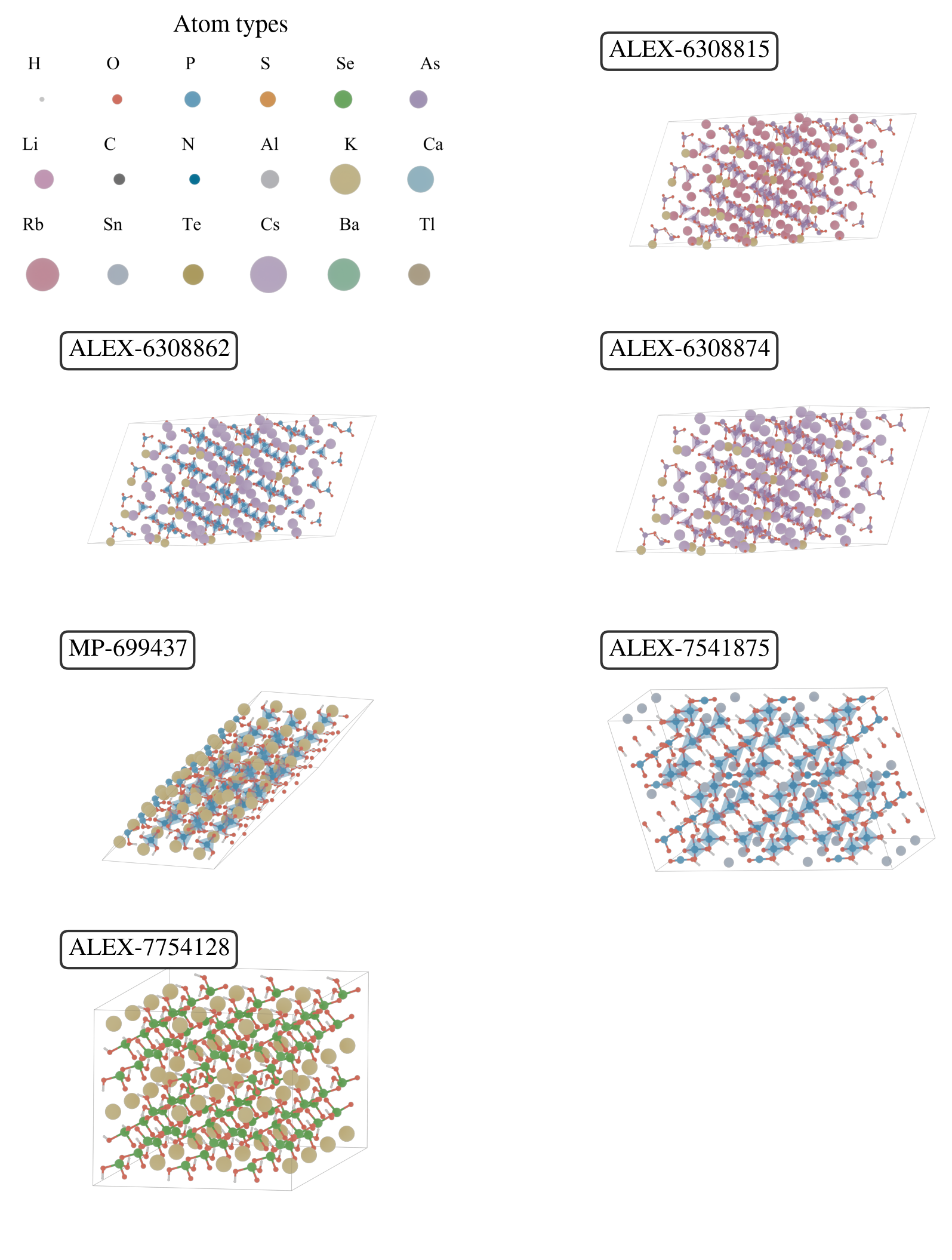}
    \caption{
    \textbf{Simulation cells of the final candidate materials.}
    Extension of Supplementary Fig.~\ref{si_fig:4}.
    }
    \label{si_fig:7}
\end{figure}

\newpage
\clearpage

\section*{Supplementary Note 6: Temperature behavior of the final candidate set}

\begin{table}[ht]
    \centering
    \caption{
    \textbf{Temperature-dependent proton transport descriptors of the final candidate set.}
    Extended version of Tab.~1 from the main text. 
    Reported are the composition, the \textsc{Materials Project} (MP-ID) or \textsc{Alexandria} (Alex-ID) identifier, as well as proton diffusion coefficients $D_\mathrm{H}$ ({\AA}$^2\,\mathrm{ps}^{-1}$), X--O bond-vector autocorrelation coefficients $C_\mathrm{XO}(\tau=0.5~\mathrm{ns})$, and proton transfer frequencies (jumps per H atom per picosecond), all obtained from $3$~ns MD trajectories at $400$~K, $500$~K, and $600$~K using material-specific \textsc{MACE-MP-0} MLIPs fine-tuned to $40$~ps AIMD reference data.
    Activation energies $E_a$ were obtained from Arrhenius fits to $D_\mathrm{H}$ across the three temperatures.
    Multiple entries for the same composition represent different polymorphs.
    All transport quantities and simulation details are described in the Methods section of the main text.
    }
    \begin{tabular}{lc@{\hspace{0.5em}}c@{\hspace{0.5em}}c@{\hspace{0.5em}}c@{\hspace{0.5em}}c@{\hspace{0.5em}}c@{\hspace{0.5em}}c@{\hspace{0.5em}}c@{\hspace{0.5em}}c@{\hspace{0.5em}}c@{\hspace{0.5em}}c@{\hspace{0.5em}}c@{\hspace{0.5em}}}
        \toprule
        & & & \multicolumn{3}{c}{$D_\mathrm{H}$\, $(\text{\AA}^2 \mathrm{ps}^{-1})$} & \multicolumn{3}{c}{$C_\mathrm{XO}(\tau=0.5~\mathrm{ns})$} & \multicolumn{3}{c}{H jumps\, ($N_\mathrm{H}^{-1}\mathrm{ps}^{-1}$)} \\
        \cmidrule(lr){4-6} \cmidrule(lr){7-9} \cmidrule(lr){10-12}
        Composition & ID & $E_a$ (eV) & 400~K & 500~K & 600~K & 400~K & 500~K & 600~K & 400~K & 500~K & 600~K \\
        \midrule
        KHSO$_{4}$ & MP-1199797 & 0.25 & 0.0048 & 0.0191 & 0.0520 & 0.155 & -0.004 & -0.001 & 15.68 & 16.85 & 16.91 \\
        (CH$_{3}$)$_{2}$NH$_{2}$H$_{2}$PO$_{4}$ & MP-1200282 & 0.25 & 0.0042 & 0.0157 & 0.0472 & 0.645 & 0.058 & -0.005 & 12.56 & 15.13 & 16.77 \\
        (NH$_{4}$)$_{3}$(HSeO$_{4}$)$_{2}$ & MP-1212963 & 0.29 & 0.0030 & 0.0172 & 0.0492 & 0.431 & 0.068 & 0.016 & 13.22 & 15.68 & 17.37 \\
        K$_{4}$LiH$_{3}$(SO$_{4}$)$_{4}$ & MP-1196827 & 0.28 & 0.0029 & 0.0122 & 0.0447 &0.296 & 0.006 & 0.001  & 11.25 & 12.58 & 13.75 \\
        H$_{56}$C$_{19}$S$_{3}$N$_{8}$O$_{15}$ & MP-1196861 & 0.37 & 0.0021 & 0.0293 & 0.0689 & 0.126 & 0.017 & 0.002 & 12.42 & 13.52 & 14.52 \\
        KH$_{2}$PO$_{4}$ & MP-699437 & 0.23 & 0.0008 & 0.0041 & 0.0076 & 0.839 & 0.511 & 0.328 & 15.32 & 15.69 & 14.59 \\
        (Al(H$_{2}$O)$_{6}$)$_{4}$H$_{4}$(SO$_{4}$)$_{7}$ & MP-1229278 & 0.28 & 0.0006 & 0.0029 & 0.0093 & 0.784 & 0.566 & 0.132 & 4.58 & 6.28 & 8.82 \\       
        Rb$_{2}$H$_{6}$(SO$_{4}$)(TeO$_{6}$) & MP-559096 & 0.21 & 0.0001 & 0.0004 & 0.0010 & 0.129 & 0.010 & -0.003 & 3.31 & 5.97 & 7.71 \\
        Tl$_{2}$H$_{6}$(SeO$_{4}$)(TeO$_{6}$) &  MP-1196566 & 0.27 & 0.0002 & 0.0011 & 0.0035 & 0.235 & -0.003 & -0.003 & 8.28 & 12.61 & 16.04 \\
        \midrule
        LiH$_{3}$(SeO$_{3}$)$_{2}$ & Alex-3230254 & 0.39 & 0.0066 & 0.1210 & 0.2571 & 0.767 & 0.043 & -0.006 & 17.34 & 15.26 & 16.57 \\
        CsHSeO$_{3}$ & Alex-7647189 & 0.26 & 0.0055 & 0.0263 & 0.0658 & 0.265 & 0.004 & 0.006 & 11.42 & 11.66 & 12.78 \\
        Sn(H$_{2}$PO$_{4}$)$_{2}$ & Alex-7541875 & 0.33 & 0.0044 & 0.0404 & 0.1063 & 0.763 & 0.133 & 0.029 & 12.65 & 14.15 & 17.11 \\
        Ca(H$_{2}$PO$_{4}$)$_{2}$ & Alex-5262379 & 0.30 & 0.0038 & 0.0195 & 0.0685 & 0.778 & 0.245 & 0.011 & 14.12 & 16.65 & 17.51 \\
        CsHSeO$_{3}$ & Alex-3230870 & 0.40 & 0.0017 & 0.0143 & 0.0862 & 0.641 & 0.082 & 0.073 & 10.61 & 11.12 & 13.54 \\
        K$_{2}$H$_{2}$As$_{2}$O$_{7}$ & Alex-6308766 & 0.29 & 0.0016 & 0.0072 & 0.0281 & 0.796 & 0.396 & 0.050 & 13.91 & 16.03 & 16.06 \\
        (KRb$_{3}$)(H$_{2}$As$_{2}$O$_{7}$)$_{2}$ & Alex-6308815 & 0.24 & 0.0016 & 0.0044 & 0.0171 & 0.784 & 0.485 & 0.116 & 15.28 & 17.11 & 15.13 \\
        (KCs$_{3}$)(H$_{2}$As$_{2}$O$_{7}$)$_{2}$ & Alex-6308874 & 0.23 & 0.0014 & 0.0034 & 0.0141 & 0.794 & 0.568 & 0.132 & 14.60 & 16.17 & 17.45 \\
        KH$_{2}$PO$_{4}$ & Alex-3240111 & 0.36 & 0.0011 & 0.0058 & 0.0406 & 0.734 & 0.192 & 0.004 & 17.26 & 16.68 & 17.38 \\
        Ba(H$_{2}$PO$_{4}$)$_{2}$ & Alex-3239126 & 0.42 & 0.0010 & 0.0108 & 0.0598 & 0.894 & 0.313 & 0.001 & 11.40 & 13.38 & 14.94 \\
        KH$_{2}$PO$_{4}$ & Alex-3240113 & 0.29 & 0.0010 & 0.0041 & 0.0165 & 0.807 & 0.448 & 0.007 & 14.46 & 16.24 & 17.01 \\
        CsHSO$_{4}$ & Alex-3236698 & 0.30 & 0.0009 & 0.0044 & 0.0163 & 0.210 & 0.067 & 0.007 & 11.20 & 11.75 & 15.65 \\
        (KCs$_{3}$)(H$_{2}$P$_{2}$O$_{7}$)$_{2}$ & Alex-6308862 & 0.23 & 0.0007 & 0.0025 & 0.0063 & 0.836 & 0.525 & 0.214 & 14.86 & 16.79 & 18.90 \\
        Rb$_{4}$H$_{4}$(Se$_{3}$S)O$_{16}$ & Alex-6308968 & 0.50 & 0.0003 & 0.0085 & 0.0332 & 0.700 & 0.076 & 0.009 & 15.17 & 17.31 & 19.43 \\
        RbHSeO$_{4}$ & Alex-6308988 & 0.50 & 0.0003 & 0.0074 & 0.0368 &  0.647 & 0.068 & -0.003 & 15.53 & 17.93 & 18.72 \\
        RbHSO$_{4}$ & Alex-5291171 & 0.49 & 0.0003 & 0.0010 & 0.0395 & 0.514 & 0.165 & 0.002 & 14.26 & 15.59 & 15.65 \\
        BaHPO$_{4}$ & Alex-3280448 & 0.12 & 0.0001 & 0.0004 & 0.0005 & 0.969 & 0.919 & 0.928 & 13.05 & 16.11 & 16.60 \\
        KH$_{3}$(SeO$_{3}$)$_{2}$ & Alex-7754128 & 0.87 & 0.0000 & 0.0319 & 0.1123 & 0.963 & 0.183 & 0.093 & 10.50 & 12.31 & 13.84 \\
        \bottomrule
    \end{tabular}
    \label{si_tab:temp}
\end{table}

\begin{figure}[ht]
    \centering
    \includegraphics{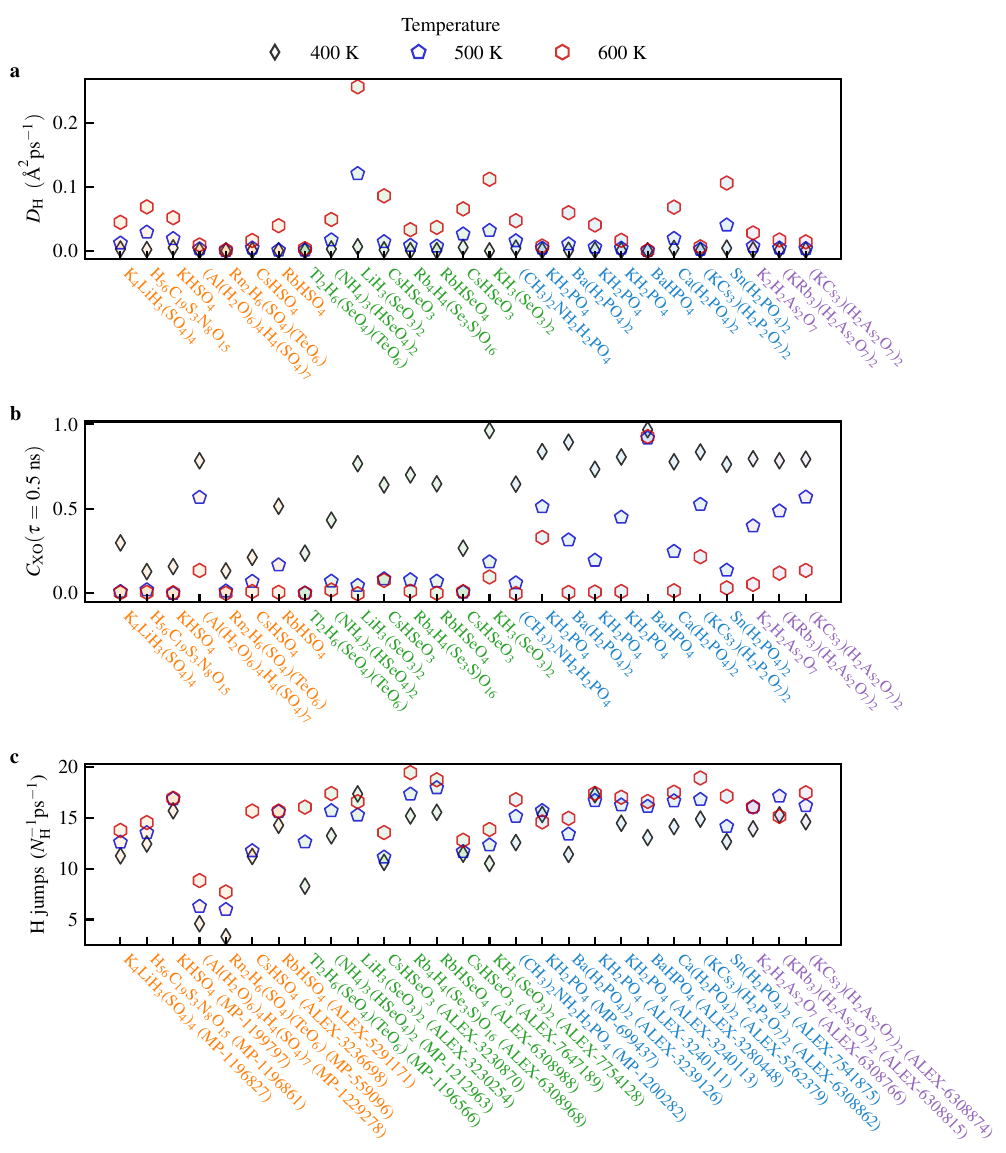}
    \caption{
    \textbf{Temperature-dependent proton transport descriptors of the final candidate set.}
    Temperature trends of the key descriptors for the final candidate set, including \textbf{a} proton diffusion coefficients, \textbf{b} oxoanion rotation dynamics, and \textbf{c} the number of proton transfer events per hydrogen atom and per picosecond.
    Values are extracted consistently from the $3$~ns fine-tuned MLIP-based MD trajectories at multiple temperatures: $400$~K, $500$~K and $600$~K, enabling a direct comparison across materials.
    All transport quantities and simulation details are described in the Methods section of the main text.
    }
    \label{si_fig:8}
\end{figure}

\newpage
\clearpage

\section*{Supplementary Note 7: Universal oxygen--oxygen distance for proton transfer in solid acid materials}

\begin{figure}[ht]
    \centering
    \includegraphics{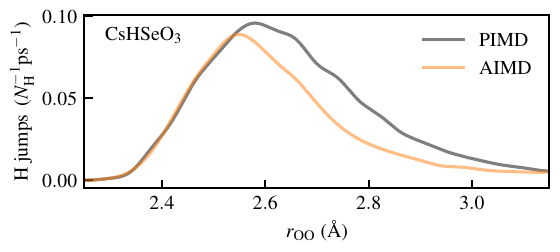}
    \caption{
    \textbf{Nuclear quantum effects and the universal oxygen--oxygen distance.}
    Distribution of proton-transfer events as a function of the instantaneous oxygen--oxygen distance, from AIMD and PIMD simulations of CsHSeO$_3$.
    Unlike TeO$_6$-containing materials (see Fig.~4b in the main text), CsHSeO$_3$ is not an outlier and shows no appreciable change in the $r_{\mathrm{OO}}$-dependence when nuclear quantum effects are included.
    PIMD simulation details are provided in the Methods section.
    }
    \label{si_fig:9}
\end{figure}

\newpage
\clearpage

\bibliography{literature_si}